\documentclass[slac_one]{revtex4}
\usepackage{graphicx}
\usepackage{fancyhdr}
\begin{document}
\newcommand{\AG}{APPLGRID}
\graphicspath{{./figures/}}

\title{
%A posteriori inclusion of parton density functions in NLO QCD calculations at hadron colliders: The APPLGrid Project
The study of uncertainties in theory predictions in single inclusive jet production cross section using the \AG\ package
} %% Paper title goes here


\author{P.~Starovoitov}
\affiliation{CERN, Geneva 23, CH-1211, Switzerland}

\begin{abstract}
A method to facilitate the consistent inclusion of cross-section measurements in proton parton density functions (PDFs) fits in NLO QCD has been developed. 
It allows the a posteriori variation of the renormalisation and factorisation scales and of the centre-of-mass energy 
and the inclusion of arbitrary PDFs, strong coupling in cross-section calculations. 
The theoretical uncertainties in the double differential single inclusive jet cross-section at a $\sqrt{s}=7\mbox{ TeV}$ have been studied using this method. 
%The technique allows to store the matrix elements for the various BSM physics scenarios calculations in the grids. 
%It is demonstrated that the method also allows to quantify the sensitivity of LHC data to deviations from SM predictions.
\end{abstract}

\maketitle

\thispagestyle{fancy}


\section{INTRODUCTION}

The Standard Model (SM) of particle physics, based on the theory of electroweak and strong interactions,  
provides a solid basis for measurements in high energy physics. The experiments at particle colliders 
LEP, HERA, TEVATRON and many others have proven the validity of the SM and measured its parameters. 
However, the SM cannot be considered as a complete theory, since it doesn't include gravity, the forth fundamental force, 
and contains too many free parameters, only few of them are measured to better than  1\% accuracy.
The new proton-proton collider LHC will produce a large amount of experimental data, which, due to extremely high luminosity,
will have small statistical uncertainty on measurements for most of the production processes. 
However, since the center of mass energy of modern accelerators is far below the Planck scale, one 
might have to look at small deviations of experimental distributions from the SM predictions to find 
signals of new particles or interactions. The understanding of the both theoretical and experimental uncertainties will be the key for a discovery.

The calculation of cross-sections at Next-to-Leading Order (NLO) in QCD involves numerical
integration over the phase space of the final state partons in order to cancel the infra-red and
collinear divergences. Since the convergence of such a calculation in certain parts of phase space might be slow, 
the integration of cross sections for observables at deep inelastic scattering (DIS) experiments or hadron-hadron 
colliders requires a time consuming generation of a large number of event weights
 in order to achieve low statistical uncertainty in theory prediction. 
Moreover, in the iterative PDF fit procedure or for the estimation of the uncertainty in 
the prediction due to PDFs, $\alpha_s$ and/or scale choice uncertainties, 
 the full calculation needs to be repeated over and over again.

Nevertheless, the factorisation of a cross section into long-distance PDFs and short-distance matrix elements (ME) 
allows to fill the latter into look-up tables (grids) binned in incoming partons momentum fraction and the characteristic scale of the production process 
during the time consuming evaluation and use it later on for the fast calculation of the theory predictions a posteriori.
Several approaches exploiting this idea \cite{Kluge:2006xs,Carli:2005ji,Carli:2010rw} have developed independently. 

The other possibility is the use of the ratio of NLO to leading order (LO) cross sections (k-factor). 
Such a factor, being calculated once, could be used to rescale the relatively fast calculation of LO cross section to mimic the NLO result,
however, this method is only an approximation.
%and should not be used in PDF fit or uncertainty estimation, 
%since k-factor internally depends on the PDF set used for its calculation.

This contribution describes the \AG\ method \cite{Carli:2010rw} and shows its usage for the evaluation 
of the theory uncertainties in the inclusive jet cross section in proton-proton collisions at 7~TeV centre-of-mass energy.

\section{OVERVIEW OF THE METHOD}

In a typical calculation of the cross section of production process $H_1H_2\rightarrow Y$ at NLO the integration over the 
process phase space is replaced by the summation 
of the ME weights {$\hat{\sigma}$} over the  set $i=1\dots N$ of kinematical configurations of partons 
factorised in the initial state hadrons with the momentum fractions { $\{x_{1},x_{2}\}_i$} at the scale  $\{Q^2_{F}\}_i$ :
 \begin{eqnarray}
 \frac{d\sigma}{dO} = \sum_{i=1}^N \sum_{\left(p_{i}\right)}    \left( \frac{\alpha_s(Q_{R_i}^2)}
 {2\pi}\right)^{p_i} \,
 f_{q_1/H_1}\left({x_1}_i,Q^2_{F_i}\right) f_{q_2/H_2}\left({x_2}_i,Q^2_{F_i}\right) 
\frac{d\hat{\sigma}^{p_i}_{q_1q_2\rightarrow Y} \left({x_1}_i,{x_2}_i,Q^2_{F_i},Q^2_{R_i}\right)}{dO}.
\label{xs} 
\end{eqnarray}

In general there are $13 \times 13$ different combinations of initial state partons contributing to an observable, however, due to the 
symmetries of the  ME weights, one can define a set of sub-processes and group the PDFs into ``generalised`` PDFs
\begin{equation}
\sum_{m,n}\nu_{mn}^{(k)}\;f_{m/H_1}\left({x_1},Q^2\right) f_{n/H_2}\left({x_2},Q^2\right) \equiv F^{(k)}\left({x_1},{x_2},Q^2\right), 
\end{equation}
where index $(k)$ denotes the sub-process.
Such ``generalised`` PDFs are, of course, depend on the process and the perturbative order. 
For example, in the case of $W-$boson production one could define two sub-processes at LO and six at NLO \cite{Carli:2010rw}.

In the beginning of the evaluation of sum (\ref{xs}) the three-dimensional $N_{x_1}\times N_{x_2}\times N_{Q^2}$ look-up 
tables in $(x_1,x_2,Q^2)$ phase space ($N_x\mbox{ and }N_{Q^2}$ are user defined) are created.
Instead of using directly the parton momentum fraction $x$ and parton resolution scale $Q^2$ 
the following variable transformation is applied
\begin{equation}
y(x) = \ln \frac{1}{x} + a(1 - x); \;\;\;
\tau(Q^2) = \ln\left(\ln\frac{Q^2}{\Lambda^2}\right).
\end{equation}
This transformation provides equidistant binning in regions of phase space, where the PDFs are steeply falling.
The parameter $\Lambda$ should be chosen of the order of $\Lambda_{\mbox{QCD}}$, but need not necessarily be identical, 
the parameter $a$ serves to increase the density of points in the large $x$ region ($\Lambda\mbox{ and }a$ are user defined).

Assumming that continious parton density function could be represented by the Lagrange interpolation 
\begin{equation}
f(x,Q^2) = \sum_{i=0}^{n_{y}} \sum_{j=0}^{n_{\tau}} f\left(y_i,\tau_j\right) 
I_i^{(n_{y})}\left(y\left(x\right)-y_i\right)
I_j^{(n_\tau)}\left( \tau\left(Q^2\right) - \tau_j \right)
\end{equation}
over $n_y$ and $n_\tau$ nodes in $(y,\tau)-$grid with sufficient precision, the sum (\ref{xs}) could be rewritten such
\begin{equation}
\label{convolution}
\frac{d\sigma}{dO}= \sum_{l=0}^L \sum_{i_{1},i_{2},j}   \left( \frac{\alpha_s(Q_{j}^2)} {2\pi}\right)^{p_{l}}
 \, F^{(l)}\left({y_{i_1}},{y_{i_2}},\tau_{j}\right) \frac{d\sigma_{(i_{1},i_{2},j)}^{l}}{dO}
\end{equation}
that interpolation is performed during the event loop over the pQCD weights rather than over PDFs. Here $Q^2_R=Q^2_F$ was assumed for simplicity, 
but this is not the fundamental limitation, since the observable scale dependence could be fully addressed a posteriori.

The \AG\ library works as an interface, which interpolates the pQCD coefficients $d\hat{\sigma}^{p_i}_{q_1q_2\rightarrow Y}$  
over  $(y_1,y_2,\tau)\equiv(x_1,x_2,Q^2)$ nodes of look-up table  and fills 
 the grid structure with one 3D-table per sub-process, perturbative order and observable bin. 

After the evaluation of the sum (\ref{xs}) the grid structure is written out to a ROOT file. The \AG\ library provides 
the methods for convolution (using the Eq. \ref{convolution}) of the pQCD coefficients stored in the grid 
with any choice of PDF set, $\alpha_s$ and scales choice.
Depending on the grid size and PDF set the convolution time is $1\div 100\mbox{ ms}$, since it is already just a simple 
sum of real numbers and not the evaluation of divergent integrals.

The choice of the grid architecture depends on the required accuracy, on the exact cross-section definition
and on the available computer resources.  The most critical parameter is the number of $x$-bins, which must be large enough
to accommodate strong PDF variations. It has been shown \cite{Carli:2010rw} that the grid reproduces the original 
calculation with the precision better than $10^{-3}$ with $N_x \approx 20 \div 30 $.

\section{SINGLE INCLUSIVE JET CROSS SECTION}

In this section we describe the use of \AG\ for the estimation of theory uncertainty  
on the example of single inclusive jet production of jets with radius $R=0.6$ for $p_{T}\ge 40\mbox{ GeV }$ 
defined by the anti-$k_{\perp}$ jet algorithm \cite{Cacciari:2008gp} in several bins of jet rapidity.
The cross section  was calculated using NLOJET++ \cite{Nagy:2003tz,Nagy:2001xb,Nagy:2001fj} with the renormalisation and factorisation scales equal to 
the $p_{T}$ of the leading jet. 

\begin{figure}[htp]
  \centering
  \includegraphics[width=.42\textwidth]{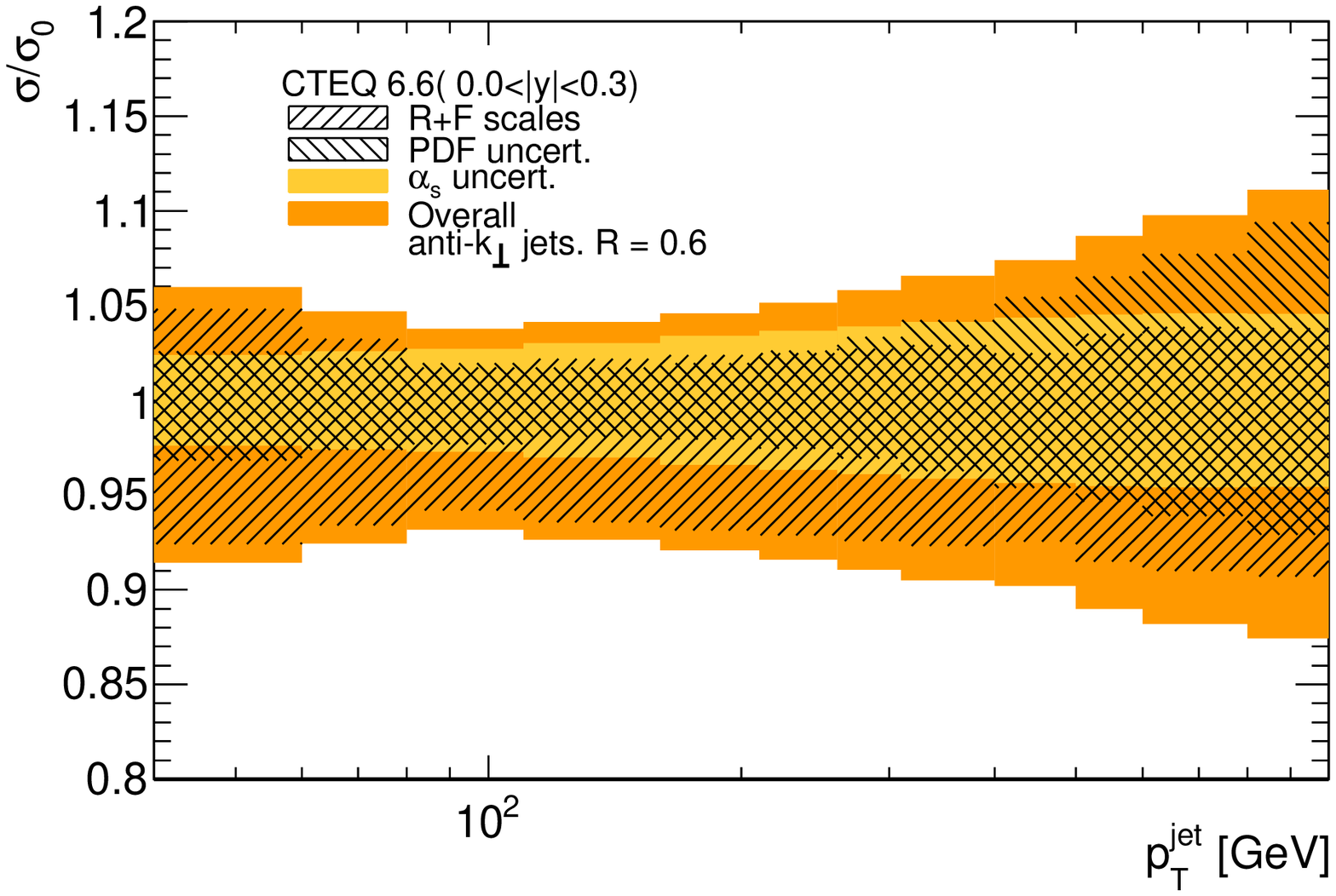}
  \includegraphics[width=.42\textwidth]{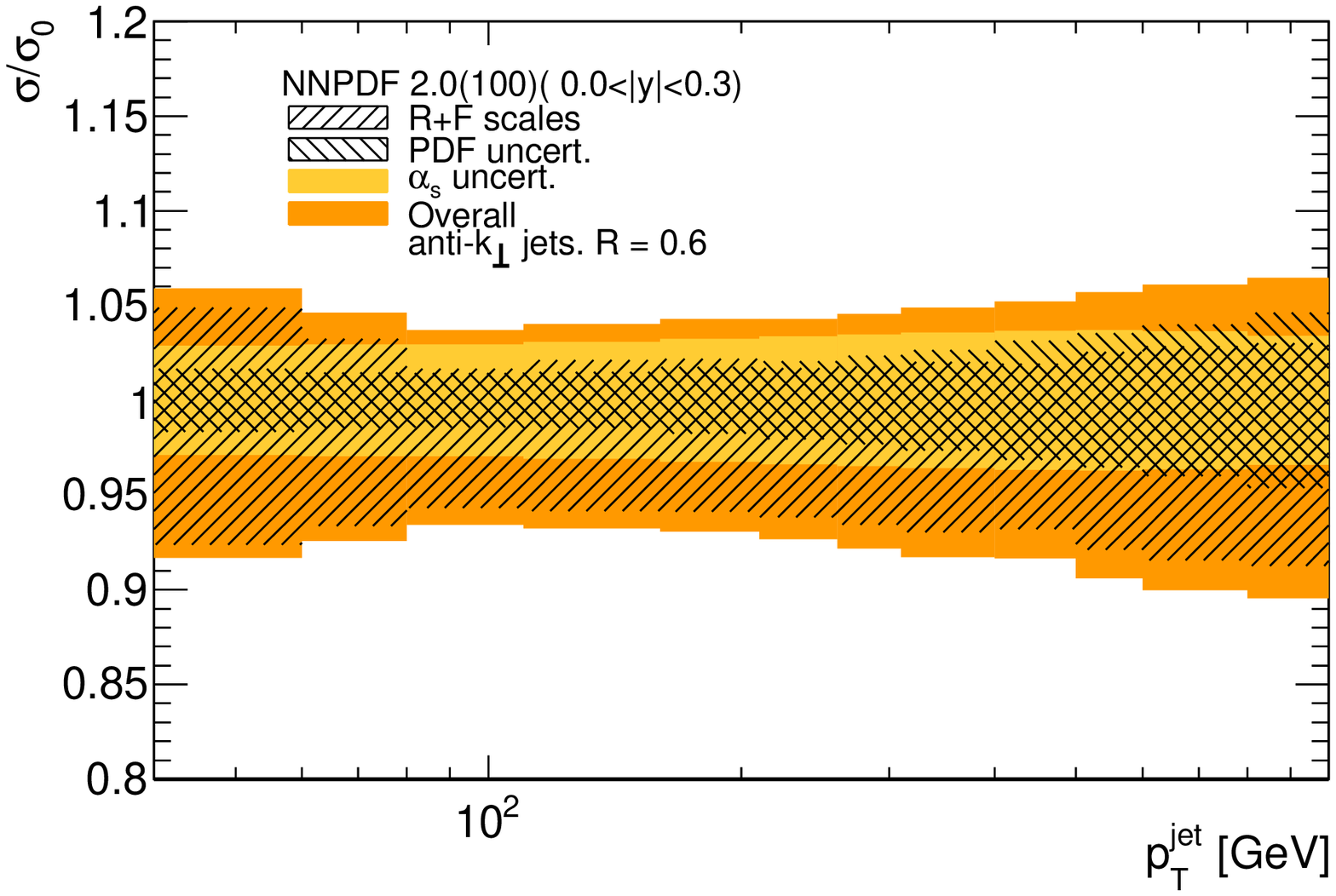}
                  \\
%\vspace*{-0.5cm}
  \includegraphics[width=.42\textwidth]{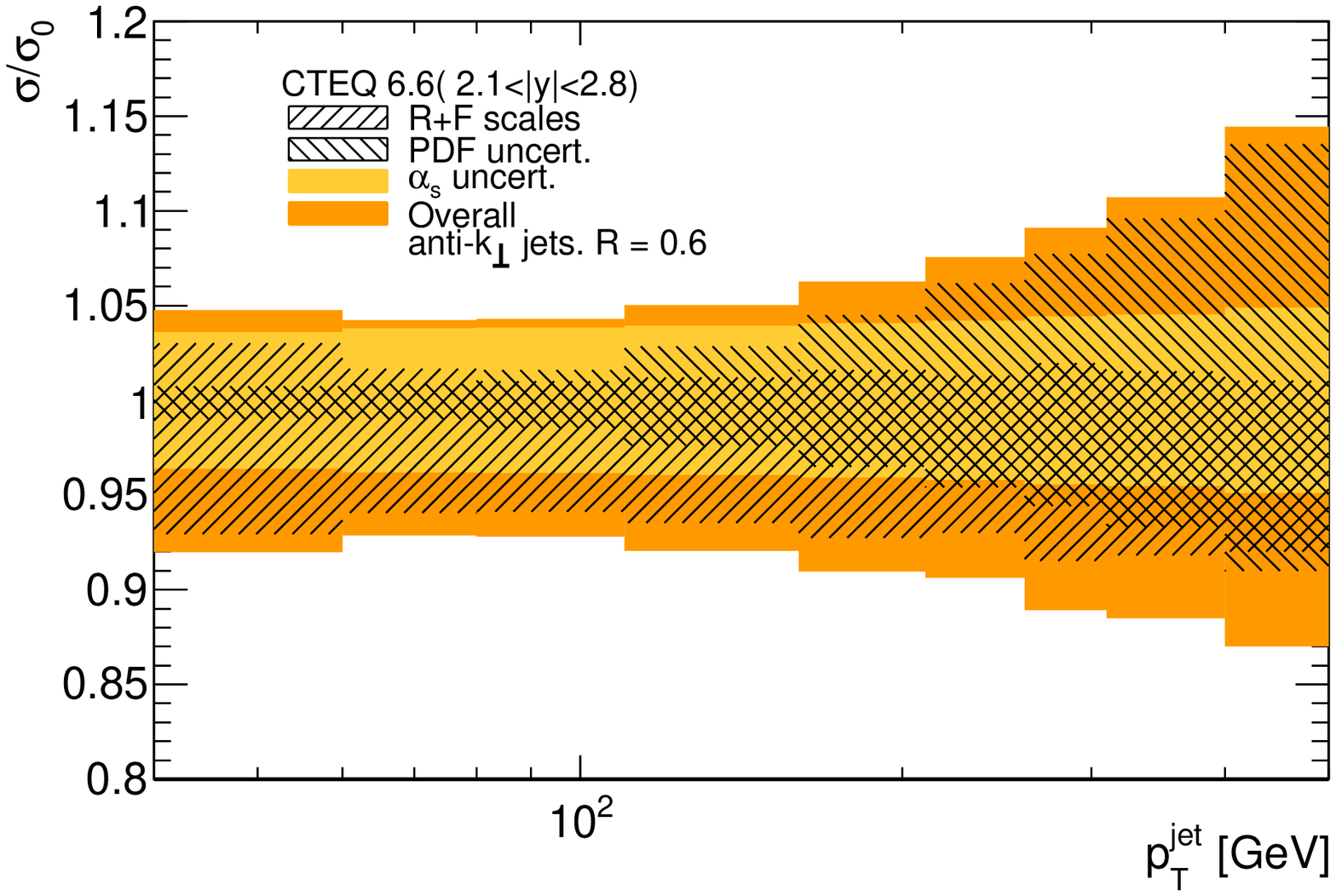}
  \includegraphics[width=.42\textwidth]{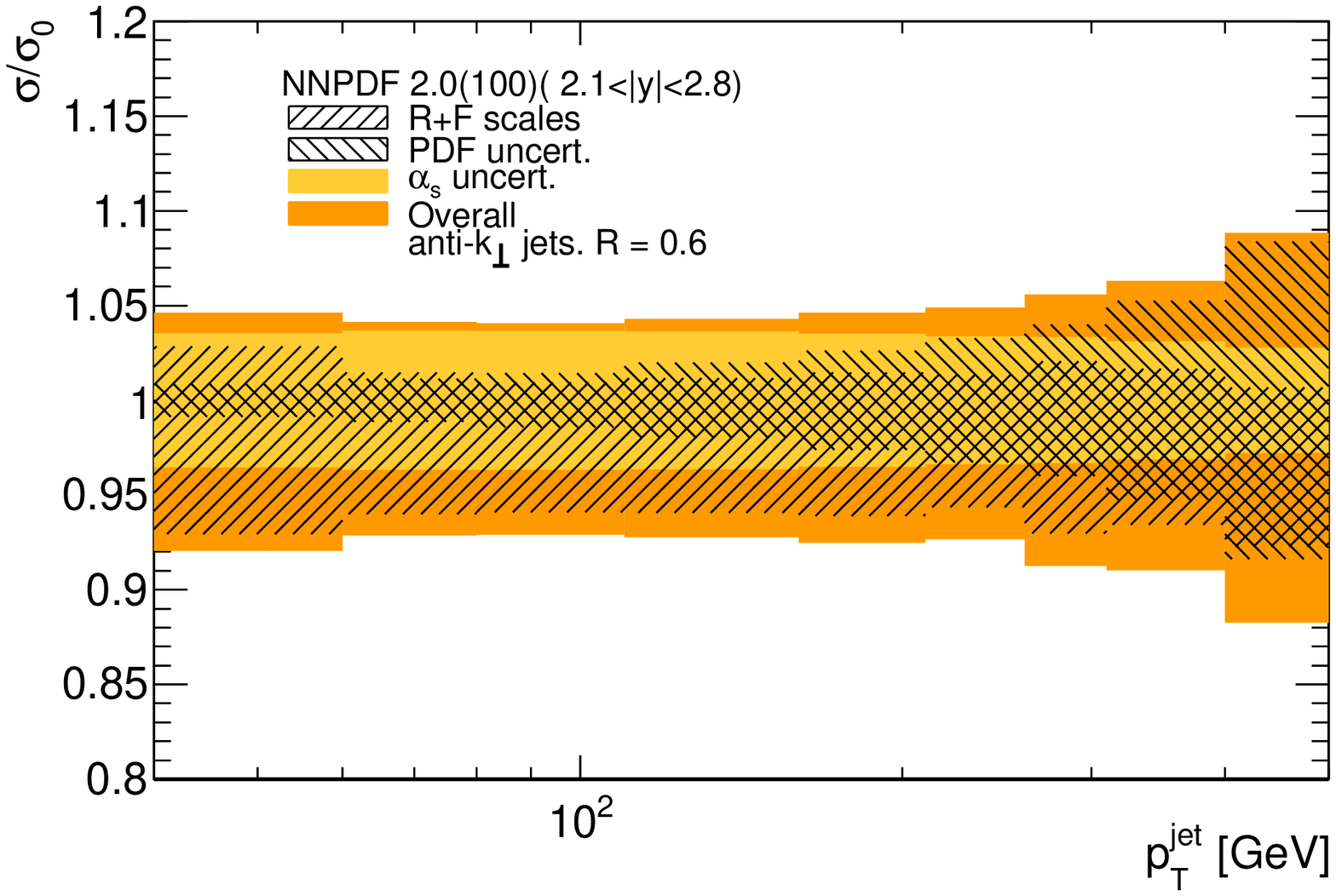}
\caption{\label{fig:singleinclusiveuncert} 
  Theory uncertainties ifor the double differential single inclusive cross section as a function of jet transverse momentum. 
The two upper (bottom) figures compare uncertainties calculated with CTEQ6.6 and NNPDF2.0  PDF sets in central (forward) jet production.
}
% \end{figure}
%
% \begin{figure}[ht]
   %\centering
%\vspace*{-0.5cm}
   \includegraphics[width=.42\textwidth]{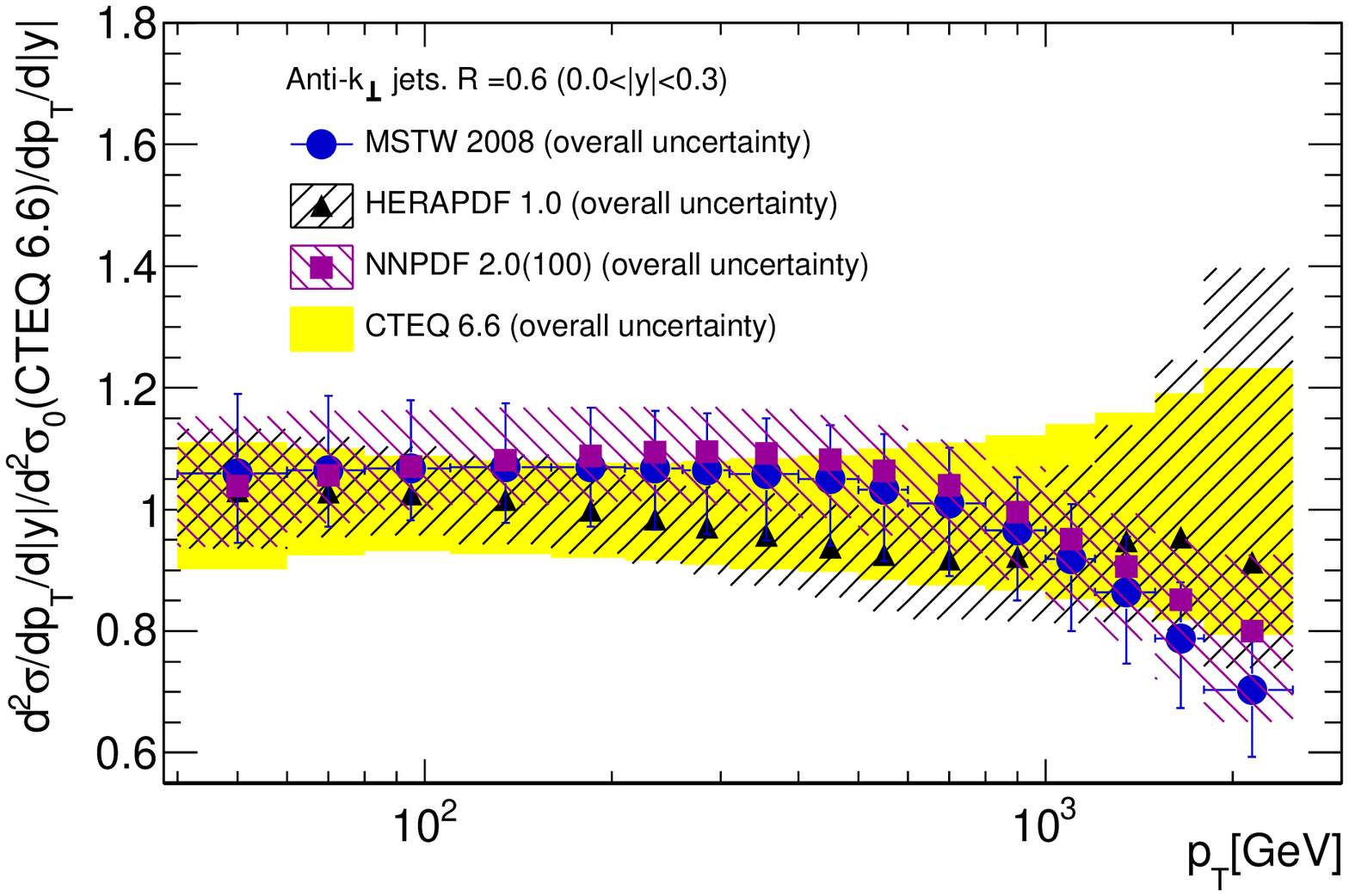}
   \includegraphics[width=.42\textwidth]{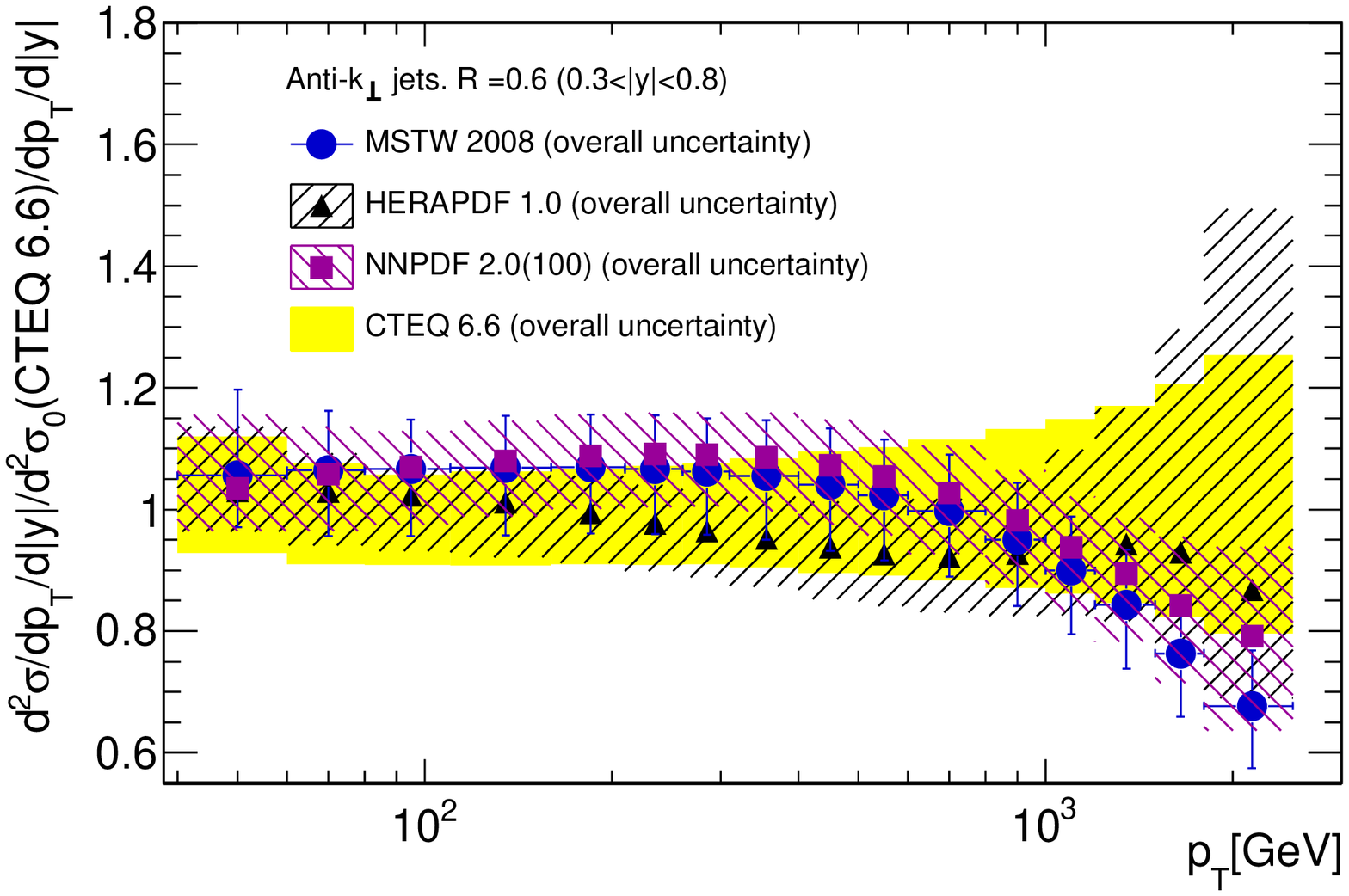}
\\
%\vspace*{-0.5cm}
  \includegraphics[width=.42\textwidth]{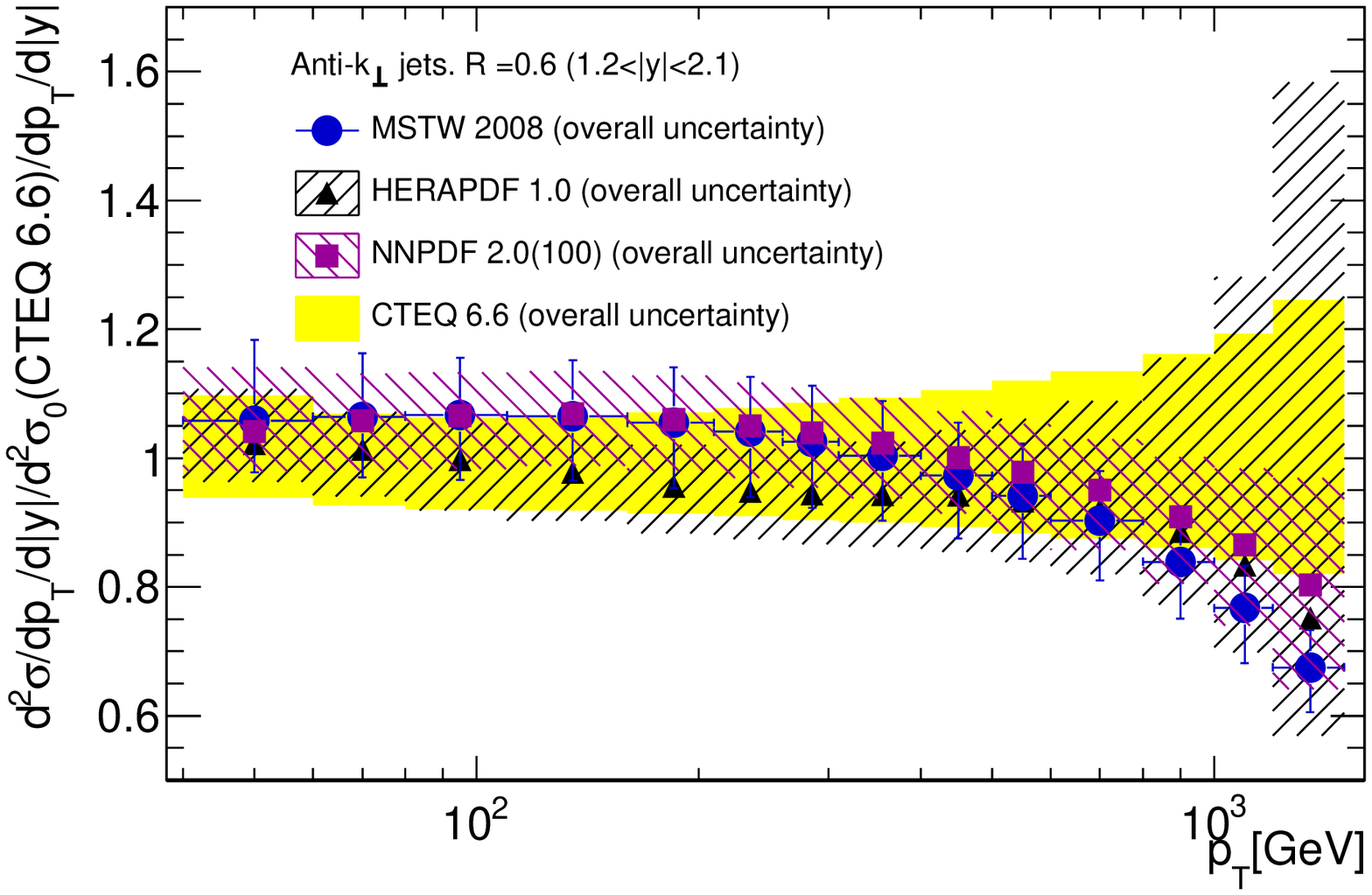}
  \includegraphics[width=.42\textwidth]{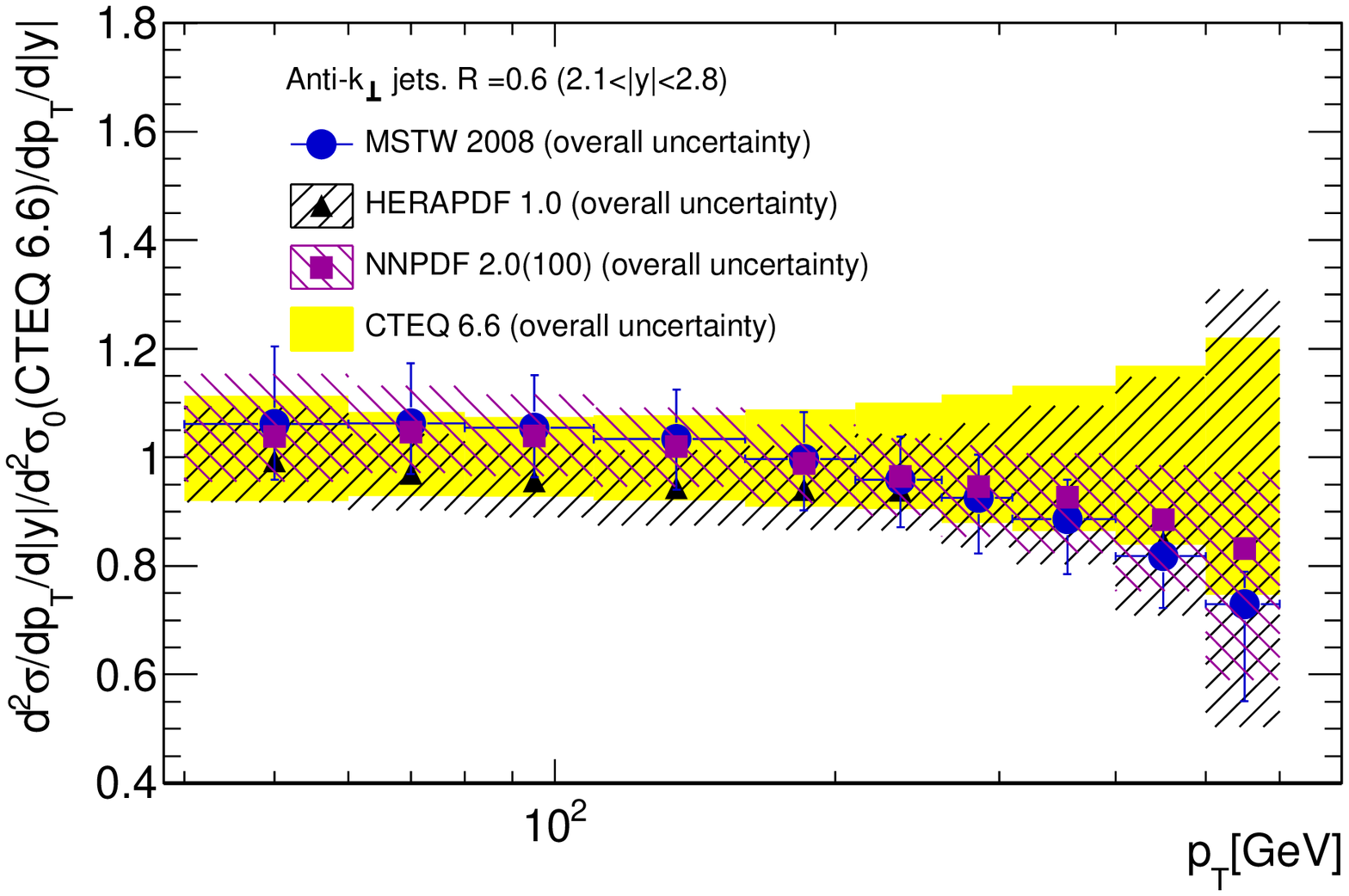}
\caption{\label{fig:singleinclusivepdfs}
Comparison of MSTW2008 \cite{Martin:2009bu} (blue circles), HEPAPDF1.0 \cite{:2009wt} (black triagnles), 
NNPDF2.0 \cite{Ball:2010de} (violet squares) and CTEQ6.6 \cite{Lai:2010nw} (yellow band) predictions 
of single inclusive jet cross section 
and the total theory uncertainty uncertainty normalised to the CTEQ6.6 central prediction 
as a function of jet transverse momentum .
}
\end{figure}

Figure \ref{fig:singleinclusiveuncert} shows the relative theory uncertainty in the cross section. 
Both the total uncertainty (orange band) and all its individual components are shown.
The scale uncertainties (right side hatches band) are defined as the envelope of the independent two scales varying 
by a factor two up and down of the default choice.
The effect of the uncertainty on the $\alpha_s$ determination (yellow band) on the observable is estimated by calculating the cross section
using  strong coupling values within the uncertainty range and using PDF sets fitted using these values \cite{Lai:2010nw}.
The PDF uncertainty (left side hatches band) is calculated using the uncertainties PDF sets as 
recommended by the authors \cite{Nadolsky:2008zw,:2009wt,Ball:2010de,Martin:2009bu}.
Finally, the individual components are added in quadrature in the total uncertainty. 
One can see that both CTEQ6.6 and NNPDF2.0 give similar uncertainty for the cross section in the central and the 
forward region for jets with $p_T < 100\mbox{ GeV.}$ 
However, the use of NNPDF2.0 allows to shrink the total theory uncertainty to $\mbox{}^{+7}_{-11}$\%  compared 
to CTEQ6.6 estimation of $\mbox{}{\pm11}$\% for high $p_T$ central rapidity jets. In the forward region the difference between 
the CTEQ6.6 and the NNPDF2.0 prediction for high $p_T$ jets is even more pronounced :  
 up to $\mbox{}^{+14}_{-13}$\% for CTEQ6.6 compared to $\mbox{}^{+9}_{-12}$\% for NNPDF2.0.
%\subsection{PDF predictions comparison in inclusive jet production}
 
Figure \ref{fig:singleinclusivepdfs} shows the comparison of different PDF sets predictions in several bins of rapidity (up to $|y|<2.8$).
Also the predictions of all the PDF sets agree between each other within the error bands, the shape of $p_T$ dependence is 
quite different for different PDF sets. All the PDFs agree within 4\% at low $(p_T<100\mbox{ GeV})$ transverse momentum, but NNPDF2.0 and MSTW2008 predict 
~5-7\% more jets at intermediate $p_{T}$'s, while HERAPDF1.0 estimates ~5\% less cross section in this $p_T$ region.
At high jet transverse momentum HERAPDF1.0, NNPDF2.0 and MSTW2008 predict, depending of rapidity region, 
 a $\sim 15-30\%$ smaller jet rate compared to CTEQ6.6.
The NNPDF2.0 and MSTW2008 provide very similar $p_T$ dependence over all the rapidity bins, they agree to CTEQ6.6 at low $p_T$, 
predict larger cross section than CTEQ6.6 at intermediate $p_T$ values and lower one for high $p_T$ jets.
The cross section calculated with HERAPDF1.0 is always smaller compared to one with CTEQ6.6, the diference is small at low $p_T$'s 
and it increases towards higher transverse momentum.

 \section{SUMMARY}

\AG\ is an open source software project written in C++  which provides a framework to create production process dependent 
look-up tables of perturbative coefficients for a posteriori calculations of observable with any given PDF set, choice of the scales and strong coupling.
The method is based on the Lagrange interpolation of PDFs and the use of symmetry in hard scattering 
allowing to group weights into sub-process contributions. %!!!!!!!!!!!!!!
It offers a robust, transparent and uniform way to study 
the theoretical uncertainties for various QCD and electroweak processes via the interfaces to jet production cross section 
calculator NLOJET++ \cite{Nagy:2003tz,Nagy:2001xb,Nagy:2001fj} and MCFM \cite{Campbell:1999ah,Campbell:2000bg}.  It provides the possibility to look at contributions 
of different sub-processes to the observable. 
\AG\ is a very efficient substitution for the use of k-factors in PDF fits to the jet and electroweak cross sections data.
The software offer additional functionality, such as a posteriori centre-of-mass energy rescaling, arbitrary variation of 
renormalisation and factorisation scales 
and the interface to use fastNLO \cite{Kluge:2006xs} grids for DIS and hadron-hadron collisions.

The use of \AG\ for a posteriori evaluation of uncertainties due to renormalisation and factorisation scale variations,
strong coupling measurements and PDFs uncertainty sets has been demonstrated on the example of single inclusive jet cross section.
The \AG\ has already been used as the standard tool by the ATLAS Collaboration for the interpretation of the jet measurements \cite{Collaboration:2010wv}.

\end{document}